\documentclass[aps,prb,showpacs,twocolumn]{revtex4}
\usepackage[dvips]{graphicx}
\usepackage{bm}
\usepackage{amssymb}
\usepackage{amsmath}

\begin{document}

\title{Band gaps in pseudopotential self-consistent $GW$ calculations}
\author{V.A. Popa, G. Brocks, and P.J. Kelly}
\affiliation{Faculty of Science and Technology and MESA$^{+}$ Research Institute,
University of Twente, P.O. Box 217, 7500 AE Enschede, The
Netherlands}
\date{\today}

\begin{abstract}
For materials which are incorrectly predicted by density functional
theory to be metallic, an iterative procedure must be adopted in order
to perform $GW$ calculations. In this paper we test two iterative
schemes based on the quasi-particle and pseudopotential approximations
for a number of inorganic semiconductors whose electronic structures
are well known from experiment. Iterating just the quasi-particle
energies yields a systematic, but modest overestimate of the band gaps,
confirming conclusions drawn earlier for CaB$_6$ and YH$_3$. Iterating
the quasi-particle wave functions as well gives rise to an imbalance
between the Hartree and Fock potentials and results in bandgaps in far
poorer agreement with experiment.
\end{abstract}


\maketitle

Materials whose metallic or semiconducting nature has not been
established experimentally pose a particular problem for
first-principles electronic structure calculations. Density
functional theory (DFT) calculations which describe them as being
metallic are not conclusive because such calculations are known to
systematically underestimate band gaps. Conventional many-body
perturbation theory (MBPT), in which the dynamically screened
Coulomb interaction is treated perturbatively, is problematic
because the size of the perturbation depends critically on whether
metallic or semiconducting screening is used. For YH$_3$ and
CaB$_6$, this problem was solved by iteration to self-consistency.
\cite{vanGelderen:prl00,Tromp:prl01} Though these particular
theoretical predictions were confirmed by subsequent experiments,
recent work casts doubt on the predictive capability of
parameter-free MBPT.
\cite{Arnaud:prb00,Ku:prl02,Kotani:ssc02,Lebegue:prb03} For this
reason, it is important to benchmark the iterative procedure by
applying it to a series of semiconducting materials whose band
gaps are well documented experimentally.

Of various many-body techniques developed in recent years to provide
corrections to DFT descriptions of the electronic structures of real
materials, the so-called $GW$ approximation,\cite{Hedin:pr65} in
which the self-energy $\Sigma$ is approximated by the product of
the one-particle Green function $G$ and the dynamically screened
electron-electron interaction $W$, has been particularly successful.
\cite{Aryasetiawan:rpp98,Hedin:jp99,Onida:rmp02}
Because $GW$ calculations are computationally very demanding, it is
common to make a number of additional approximations. The Green
function is usually approximated by a quasi-particle (QP) expression,
where the QP energies are interpreted as a band structure. A further
approximation involves using an independent particle expression for
$G$ in the computational steps required to evaluate $W$ and $\Sigma$.
Generally the energy levels and orbitals obtained in a DFT calculation
are used in this expression, most often at the local density
approximation (LDA) level.\cite{Hohenberg:PR64,Kohn:PR65}
If additionally the QP wave functions are approximated by the LDA
orbitals, then the QP energies simply become first-order perturbation
corrections to the LDA energy levels. A $GW$ calculation that involves
all these approximations is referred to as ``single shot''. The single
shot $GW$ approach yields good QP band structures and reasonable band
gaps for a wide range
\cite{Aryasetiawan:rpp98,Hedin:jp99,Onida:rmp02} of semiconductors and
insulators provided the reference independent particle spectrum, i.e.,
the LDA spectrum, has a band gap.
\cite{Hybertsen:prl85,Hybertsen:prb86,Godby:prl86,Godby:prb88}

For materials such as Ge or InAs which LDA \textit{wrongly} predicts
to be nearly metallic or metallic, the single shot $GW$ procedure is
no longer appropriate. Iteration was suggested as a solution to this
problem,\cite{Hybertsen:prb86} but in practice the computational
expense was prohibitive and instead use was made of the experimentally
known dielectric screening to study such materials \cite{Zhu:prb91}
thereby introducing an element of empiricism. For materials whose
experimental character is unclear but which are described by the LDA
as being metallic, there is no other option than to iterate the $GW$
solution. Though this iterative $GW$ technique has been applied
successfully to YH$_{3}$ with 24 atoms in its primitive unit cell and
to CaB$_{6}$, its general validity has yet to be demonstrated. We will
apply it to a set of materials with band gaps that range from
0.4 to 5.5 eV: InAs, Ge, Si, GaAs, AlP, GaN and C. Note, in particular,
that the small band gap semiconductor InAs has a metallic LDA spectrum
and that Ge has a vanishingly small LDA band gap.

\textit{Computational details.} Our main interest lies in developing a practical scheme to
handle complex materials, such as YH$_{3}$, with large unit cells, and we have not attempted
to go beyond the QP approximation. \cite{Schone:prl98,Ku:prl02} For the same reason we use
pseudopotentials (PP) to represent the ion cores, since for complex materials all-electron
$GW$ calculations are computationally even more demanding.
\cite{Arnaud:prb00,Kotani:ssc02,Tiago:prb04,Lebegue:prb03} We use the real-space
imaginary-time $GW$ technique\cite{Rojas:prl95,Rieger:cpc99} as implemented by van Gelderen
\textit{et al.}\cite{vanGelderen:prb02} Our starting point is an LDA calculation with a
standard exchange-correlation functional, \cite{Ceperley:prl80,Perdew:prb81} normconserving
pseudopotentials \cite{Troullier:prb91} and an expansion of the orbitals in a plane wave
basis set. The LDA eigenvalues and orbitals are used to set up an initial Green function $G$.
The dynamically screened electron-electron interaction $W$ is then calculated using the
random phase approximation (RPA) and from $G$ and $W$ the self-energy $\Sigma$ is
constructed. Within the QP approximation the fundamental equation is
\begin{eqnarray}
\left[-\frac{1}{2}\nabla^{2}+v_{\mathrm{ext}}(\mathbf{r})+V_{H}(\mathbf{r})\right]
                                                       \psi_{n\mathbf{k}}(\mathbf{r})
&+& \int d\mathbf{r}'\Sigma^{x}(\mathbf{r},\mathbf{r}')\psi_{n\mathbf{k}}(\mathbf{r}')  \nonumber \\
  + \int d\mathbf{r}'\Sigma^{c}(\mathbf{r},\mathbf{r}';\varepsilon_{n\mathbf{k}})
                                                       \psi_{n\mathbf{k}}(\mathbf{r}')
&=&                          \varepsilon_{n\mathbf{k}} \psi_{n\mathbf{k}}(\mathbf{r}),
\label{QP_eq}
\end{eqnarray}
where $v_{\mathrm{ext}}$ stands for the sum of all ionic pseudopotentials,
$V_{H}$ is the Hartree potential and
$\Sigma^{x}$ is the exchange or Fock potential.
$\Sigma^{c}$ is the correlation part of the self-energy, which includes
the dynamic screening and is energy dependent.
$\varepsilon_{n\mathbf{k}}$ and $\psi_{n\mathbf{k}}(\mathbf{r})$ are,
respectively, the QP energy and wave function.
The $\varepsilon_{n\mathbf{k}}$ are complex; the real parts comprise
the band structure and the imaginary parts give the inverse QP lifetimes.

In practice \eqref{QP_eq} is solved by expanding the QP wave
functions in the orthonormal basis formed by the LDA orbitals
$\psi_{n\mathbf{k}}(\mathbf{r})
 =\sum_{n'}c_{n',n\mathbf{k}}\psi_{n'\mathbf{k}}^{\mathrm{LDA}}(\mathbf{r})$.
Representing the potentials and self-energy on the same basis,
\eqref{QP_eq} is transformed into a matrix eigenvalue equation,
\begin{equation}
\left[\mathbf{h}_\mathbf{k}+\mathbf{\Sigma}_\mathbf{k}^c(\varepsilon_{n\mathbf{k}})\right]
\mathbf{c}_{n\mathbf{k}}=\varepsilon_{n\mathbf{k}}\mathbf{c}_{n\mathbf{k}},
\label{mat_eq}
\end{equation}
where $(\mathbf{h}_\mathbf{k})_{n'n}=\langle
\psi_{n'\mathbf{k}}^{\mathrm{LDA}}|-\frac{1}{2}\nabla^{2}+v_{\mathrm{ext}}+V_{H}+\Sigma^x|
\psi_{n\mathbf{k}}^{\mathrm{LDA}}\rangle$ and $(\mathbf{\Sigma}_\mathbf{k}^c)_{n'n}=\langle
\psi_{n'\mathbf{k}}^{\mathrm{LDA}}|\Sigma^c|\psi_{n\mathbf{k}}^{\mathrm{LDA}}\rangle$. This
equation is non-linear in $\varepsilon_{n\mathbf{k}}$ because
of the energy dependence of $\Sigma^{c}$. It is solved by standard
diagonalization and root-searching techniques. If the QP and LDA
wave functions do not differ significantly, so that
$\psi_{n\mathbf{k}}(\mathbf{r})\approxeq\psi_{n\mathbf{k}}^{\mathrm{LDA}}(\mathbf{r})$
or $c_{n',n\mathbf{k}}=\delta_{n'n}$, then \eqref{mat_eq} breaks
up into a set of decoupled equations that involve the diagonal
matrix elements only\cite{Hybertsen:prb86,Godby:prb88}
\begin{equation}
(\mathbf{h}_\mathbf{k})_{nn}+(\mathbf{\Sigma}_\mathbf{k}^c(\varepsilon_{n\mathbf{k}}))_{nn}
=\varepsilon_{n\mathbf{k}}. \label{dec_eq}
\end{equation}

In our iterative procedures the QP energies and, optionally, the QP
wave functions are used to construct a new Green function $G$. We
approximate $G$ by an independent particle Green function, using only
the real part of the QP energies and setting the QP weights to unity
in order to obey the sum rule for the spectral density. The QP wave
functions are then explicitly orthogonalized. After constructing a new
$W$ and $\Sigma$, the QP equation is solved again. This procedure is
iterated until there are negligible changes in the QP energies. Since
the first iteration starts from the LDA spectrum, it can sometimes be
convenient to open a band gap artificially in this spectrum. We checked
that the converged band gap in the QP spectrum does not depend on the
size of the initial band gap.

For the materials treated in this paper, all iteration schemes to be
discussed below converge to 1 meV or better within 9 iterations.
Spin-orbit coupling is not taken into account explicitly in the solids,
but the end results are corrected a posteriori when necessary. Typical
values for the parameters used in our $GW$ calculations are a real
space mesh of (6$\times$6$\times$6) points in the unit cell,
(8$\times$8$\times$8) unit cells in the interaction cell and 200
(occupied plus unoccupied) bands.
\cite{Rojas:prl95,Rieger:cpc99,vanGelderen:prb02} The results are
extrapolated to absolute convergence in the individual parameters,
and the absolute values of these corrections are summed to define
a computational error bar.

\textit{Iteration on QP energies.} The simplest iteration scheme involves QP energies only.
The QP wave functions are assumed to be identical to the LDA orbitals and \eqref{dec_eq} is
solved to obtain $\varepsilon_{n\mathbf{k}}$. These are used to update $G$, $W$, and
$\Sigma^{c}$ before \eqref{dec_eq} is solved again and the procedure repeated until the
$\varepsilon_{n\mathbf{k}}$ are converged. Note that between iterations only the correlation
part of the self-energy needs to be updated.

In the case of InAs care must be taken, since in the LDA band structure
the valence and conduction bands overlap. In particular, the
``light hole'' valence band mixes with the lowest conduction band
leading to an avoided crossing around $\Gamma$. In our $GW$ calculations
we allow these two bands to ``de-mix'' by calculating the off-diagonal
matrix element of the self-energy in \eqref{mat_eq} and solving a
$2\times 2$ version of \eqref{mat_eq} instead of \eqref{dec_eq}.

\begin{figure}[btp]
\includegraphics[width=8cm]{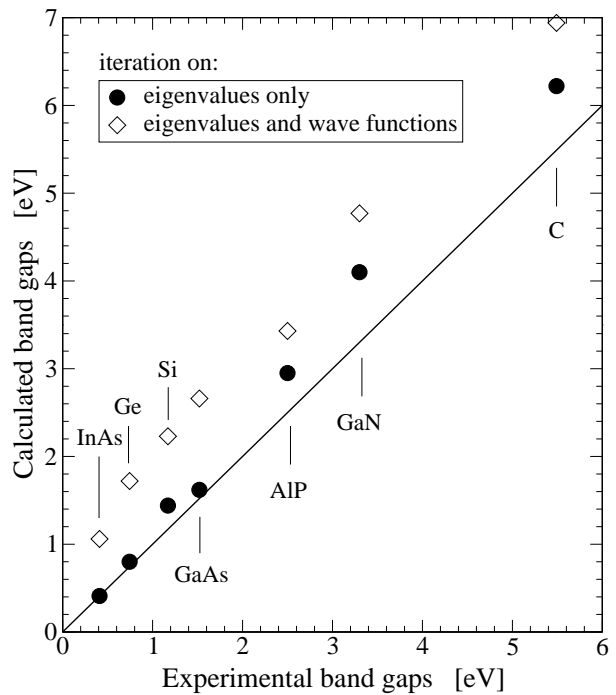}
\caption{Comparison of calculated and experimental band gaps for 7
representative inorganic semiconductors using two different
iterative $GW$ schemes discussed in the text.} \label{fig}
\end{figure}

The band gaps calculated in this way (``eigenvalues only'') are compared
with experiment in Fig.~\ref{fig} and the numerical values are given in
the third column of Table~\ref{tab}. They are seen to overestimate the
experimental values by 5 to 25\%. Assuming such overestimates also hold
for YH$_{3}$ and CaB$_{6}$ whose band gaps were calculated using the same
approximations,\cite{vanGelderen:prl00,Tromp:prl01,vanGelderen:prb02}
we arrive at an uncertainty in the predicted band gaps of
$\lesssim 0.2$ eV, which does not change the prediction that these
compounds are semiconductors. The
advantage of the iteration procedure is that the final, self-consistent
gap does not depend upon whether the initial LDA band structure corresponds
to a metal or a semiconductor.

\begin{table}[btp]
\caption{Band gaps of 7 representative inorganic semiconductors.
The experimental values are taken from Ref.[\onlinecite{Hellwege:82}],
except when stated otherwise.
The theoretical band gaps were calculated using the two different
iterative $GW$ schemes described in the text and have had a spin-orbit
splitting correction added to facilitate comparison with experiment.
The theoretical ``error-bars'' are our best estimate of the
uncertainties which result from using a finite cutoff, k-point
sampling etc.
All values are in eV.}
\label{tab}
\begin{ruledtabular}
\begin{tabular}{llcc}
     & Exp.   & QP energies  & QP energies + functions  \\
\colrule
InAs &  0.41        &  0.41 $\pm$ 0.03  &  1.06 $\pm$ 0.10  \\
Ge   &  0.74        &  0.80 $\pm$ 0.03  &  1.72 $\pm$ 0.11  \\
Si   &  1.17        &  1.44 $\pm$ 0.04  &  2.23 $\pm$ 0.09  \\
GaAs &  1.52        &  1.62 $\pm$ 0.05  &  2.66 $\pm$ 0.15  \\
AlP  &  2.50        &  2.95 $\pm$ 0.03  &  3.43 $\pm$ 0.12  \\
GaN  &  3.30$^{a}$  &  4.10 $\pm$ 0.05  &  4.77 $\pm$ 0.31  \\
 C   &  5.49        &  6.22 $\pm$ 0.04  &  6.94 $\pm$ 0.17  \\
\end{tabular}
\end{ruledtabular}
\footnotetext{$^{a}$Reference \onlinecite{Sitar:JMSL92}.}
\end{table}

\begin{figure}
\includegraphics[width=8cm]{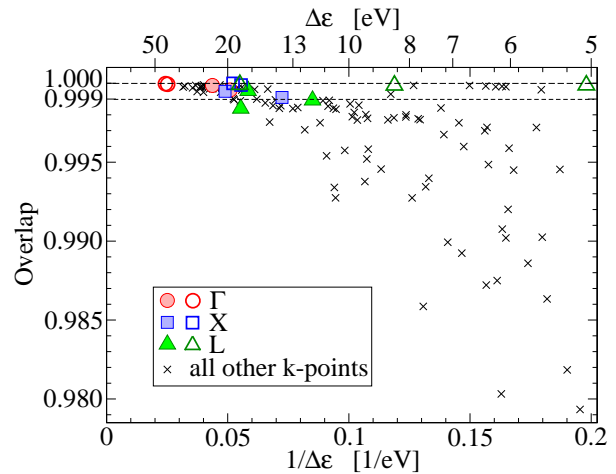}
\caption{ (Color online) Overlap of the QP wavefunction with its LDA counterpart, $
\left|\langle \psi_{n\mathbf{k}}^{\mathrm{LDA}}|\psi_{n\mathbf{k}} \rangle \right|$, plotted
as a function of $1/{\Delta \varepsilon}$, the inverse separation to the LDA state with the
next largest overlap, for the 4 lowest QP states of silicon and 39 irreducible k-points. At
the symmetry points $\Gamma$, X and L the four occupied states are shown as shaded/coloured
symbols and, in addition, the four lowest unoccupied states are shown as open symbols.}
\label{fig1}
\end{figure}

\textit{Iteration on QP energies and wave functions.}
So far we have used ``self-consistency'' to refer to the QP energies
only, and have approximated the QP wave functions by LDA orbitals.
While it is generally believed that the QP wave functions
$\psi_{n\mathbf{k}}$ are virtually identical to the LDA orbitals,
with overlaps
$ \left|\langle \psi_{n\mathbf{k}}^{\mathrm{LDA}}|\psi_{n\mathbf{k}} \rangle \right|$
of 99.9\%,\cite{Hybertsen:prb86,Godby:prb88} there have been some
suggestions that the differences, though small, might be important.
\cite{Schone:prl98}
By solving the full matrix equation \eqref{mat_eq} in a single iteration,
we confirm previous results for large overlap values but find that they
only hold for points and lines of high symmetry. For points of lower
symmetry, substantial deviations from unity are found (Fig.~\ref{fig1}).
When the QP wave functions as well as the QP energies are updated the
deviation becomes even larger. The reason why the overlaps are so
large at points of high symmetry is quite simple: the higher the
symmetry of a ${\bf k}$-point, the greater the number of irreducible
representations, resulting in a larger average separation of levels
which can be coupled by the self-energy.
Because the overlaps are larger than expected, we also investigate the
effects of iterating the QP wave functions in a self-consistent procedure
involving the solution of the full\cite{fn1:band_matrix} matrix equation
\eqref{mat_eq}.

The band gaps calculated by iterating the QP energies and wave functions
are given in the fourth column of Table \ref{tab} and are also shown in
Fig.~\ref{fig}. They overestimate\cite{Schone:prl98} the experimental
values by between 1.0 and 1.3 eV. Iterating the wave functions thus
has a very significant effect on the energy spectra, even though the
final QP wave functions still have a large overlap (mean overlap of
more than 90\%) with the LDA wave functions. Relatively small changes
in the wave functions result in large changes in the self-energy and
in the band gaps.

It turns out that the Hartree and the Fock potentials,
\begin{equation}\label{eq_Vh}
  V_{H}(\mathbf{r})=e^{2}\sum_{n\mathbf{k}}^{occ}\int d\mathbf{r}'
  \frac{|\psi_{n\mathbf{k}}(\mathbf{r}')|^{2}}{|\mathbf{r}-\mathbf{r}'|}
\end{equation}
respectively,
\begin{equation}\label{eq_Vx}
  \Sigma^{x}(\mathbf{r},\mathbf{r}')=-e^{2}\sum_{n\mathbf{k}}^{occ}
  \frac{\psi_{n\mathbf{k}}(\mathbf{r})\psi_{n\mathbf{k}}^{\ast}(\mathbf{r}')}{|\mathbf{r}-\mathbf{r}'|}
\end{equation}
are especially sensitive to the wave functions.
Both potentials are large, so small changes in the wave functions can
have a large absolute effect. This is confirmed by fixing $V_{H}$ at
the LDA level while iterating $\Sigma^{x}$ and $\Sigma^{c}$ using the
QP energies and (orthogonalized) wave functions. This leads to a band
gap of 4.35 eV for silicon ! The Fock potential $\Sigma^{x}$ is
attractive and during iteration its effect on the occupied states
increases, whereas the effect on the unoccupied states decreases,
which enlarges the band gap. This effect is partially cancelled by
iterating the repulsive Hartree potential $V_{H}$, which behaves
similarly to $\Sigma^{x}$ upon iteration, but with an opposite effect
on the band gap due to its sign. The cancellation however is
incomplete resulting in band gaps that are considerably too large
as seen in Table~\ref{tab} and Fig.~\ref{fig}.

It is very tempting to speculate that this cancellation would be more
efficient in an all-electron (AE) calculation. A number of single-shot
AE calculations yield values for the band gap of silicon which are too
small compared to experiment;\cite{Ku:prl02,Kotani:ssc02,Lebegue:prb03}
see the third column of Table~\ref{tab2}. In an even more recent paper
reporting on the iteration of an all-electron GW scheme to
self-consistency, a value for the bandgap of Si of 1.14 eV was found.%
\cite{Faleev:prl04}

Because AE GW calculations are much more expensive than PP
calculations, it is important to understand the source of the
discrepancy between the two in order to develop computationally
less demanding approximate schemes. If we accept that an
all-electron, iterative GW scheme yields results in agreement with
experiment then we must conclude that the good agreement with
experiment found by single-shot plane-wave pseudopotential (PP)
calculations is accidental and comes about from a cancellation of
two errors. Since the
single-shot AE calculations%
\cite{Ku:prl02,Kotani:ssc02,Lebegue:prb03,Faleev:prl04} for Si find
band gaps ranging from 0.84 to 0.92 eV, the PP scheme would appear to
introduce an error of 0.25-0.33 eV.\cite{fn2:Tiago}
Because the first-principles pseudopotentials
are constructed so as to yield the same Kohn-Sham eigenvalues spectrum
as the AE calculation, the error must be introduced by the
pseudowavefunctions in the polarization function where it enters via
the dipole matrix elements; this could be confirmed by explicit
comparison of AE and PP calculations.
The error in the single-shot AE schemes are apparently almost exactly
compensated on iteration.\cite{Ku:prl02,Faleev:prl04} In the PP schemes,
the combination of iteration and pseudo-wavefunctions leads to an
amplification of the error.\cite{fn3:Schoene}

\begin{table}[btp]
\caption{Comparison of the results of a number of pseudopotential (PP) and all-electron (AE)
$GW$ calculations for the fundamental band gap of Si, based upon the Dyson equation (DE), or
the quasi-particle approximation (QP). All values are in eV.} \label{tab2}
\begin{ruledtabular}
\begin{tabular}{llcc}
Reference                                & Method                    & $\Sigma_{GW}(G_{LDA}) $
                                                                              & $\Sigma_{GW}(G_{GW})$ \\
\colrule
\protect{ [\onlinecite{Schone:prl98}] }  & PP/DE  26 lowest states   &  1.34  &  1.91  \\
\protect{ [\onlinecite{Ku:prl02}]     }  & AE/DE; $\Sigma $ diagonal &  0.85  &  1.03  \\
\protect{ [\onlinecite{Kotani:ssc02}] }  & AE/QP                     &  0.90  &        \\
\protect{ [\onlinecite{Faleev:prl04}] }  & AE/QP                     &  0.84  &  1.14  \\
\protect{ [\onlinecite{Lebegue:prb03}] } & AE/QP                     &  0.92  &        \\
\hline
Present                                  & PP/QP                     &  1.17  &  2.23  \\
Present                                  & PP/QP; 26 lowest states   &  1.04  &  1.94  \\
\hline
\protect{ [\onlinecite{Hellwege:82}] }   & Experiment                & \multicolumn{2}{c} {1.17}  \\
\end{tabular}
\end{ruledtabular}
\end{table}

\textit{Conclusions.} We have tested two iterative, self-consistent $GW$ schemes within the
quasi-particle approximation, treating only the valence electrons and using pseudopotentials,
on a range of semiconductors and insulators. Iterating the QP energies to self-consistency
leads to an overestimation of the band gaps by 5 to 25\%, which is sufficiently accurate for
this scheme to have predictive power and leaves unchanged the conclusions drawn in Refs.
[\onlinecite{vanGelderen:prl00,Tromp:prl01}]. If the QP wave functions are allowed to change
in the iterations, the results get worse and the band gaps are overestimated by up to 1.3 eV.
More work needs to be done to understand in detail the origin of this error.


\begin{thebibliography}{31}
\bibitem[{fn1()}]{fn1:band_matrix}
\bibinfo{note}{To reduce the computational effort somewhat we observe that the
  contribution of the off-diagonal matrix elements decreases with their
  separation in energy from the matrix diagonal. We therefore approximate the
  full $\mathbf{h}_\mathbf{k}$ and $\mathbf{\Sigma}_\mathbf{k}^c$ matrices by
  band diagonal form. In our case the dimension of the matrices is 200. Using
  only 40 off-diagonal elements on each side of the diagonal gives converged QP
  energies that are within 4 meV of the values obtained if full matrices are
  used.}
\bibitem[{fn3()}]{fn3:Schoene}
\bibinfo{note}{We can reconcile the value of 1.91 found by Sch\protect{\"{o}}ne
  and Eguiluz [\onlinecite{Schone:prl98}] with our own value of 2.23 eV by
  iterating a PP GW calculation including only 26 states; we then find
  agreement with these authors within the error bar of the calculation. We have
  no explanation for their finding a single-shot Si band gap of 1.34 eV. Our
  own value of 1.17 (1.04 with 26 states) eV is compatible with other
  single-shot PP-GW studies.}
\end{thebibliography}

\end{document}